\documentclass{Interspeech}

\interspeechcameraready

\title{Can Quantized Audio Language Models Perform Zero-Shot Spoofing Detection?}

\author[]{Bikash}{Dutta*}
\author[]{Rishabh}{Ranjan*}
\author[]{Shyam}{Sathvik}
\author[]{Mayank}{Vatsa}
\author[]{Richa}{Singh}

\affiliation{}{Indian Institute of Technology Jodhpur}{India}
\email{\{d22cs051, ranjan.4, b22ee036, mvatsa, richa\}@iitj.ac.in}

\usepackage{comment}
\usepackage{multirow}

\begin{document}
\maketitle
\begin{abstract}


Quantization is essential for deploying large audio language models (LALMs) efficiently in resource-constrained environments. However, its impact on complex tasks, such as zero-shot audio spoofing detection, remains underexplored. This study evaluates the zero-shot capabilities of five LALMs, GAMA, LTU-AS, MERaLiON, Qwen-Audio, and SALMONN, across three distinct datasets: ASVspoof2019, In-the-Wild, and WaveFake, and investigates their robustness to quantization (FP32, FP16, INT8). Despite high initial spoof detection accuracy, our analysis demonstrates severe predictive biases toward spoof classification across all models, rendering their practical performance equivalent to random classification. Interestingly, quantization to FP16 precision resulted in negligible performance degradation compared to FP32, effectively halving memory and computational requirements without materially impacting accuracy. However, INT8 quantization intensified model biases, significantly degrading balanced accuracy. These findings highlight critical architectural limitations and emphasize FP16 quantization as an optimal trade-off, providing guidelines for practical deployment and future model refinement.

\end{abstract}
\def\thefootnote{*}\footnotetext{These authors contributed equally to this work}\def\thefootnote{\arabic{footnote}}

\section{Introduction}

Large Audio Language Models (LALMs) \cite{peng2024survey} have significantly advanced audio processing tasks such as speech recognition, sound event detection, and audio captioning \cite{NEURIPS2023_3a2e5889, gong2023listen}. These models, trained on extensive multimodal datasets, frequently achieve near-human proficiency, demonstrating impressive capabilities in complex audio understanding and generation. A key feature of LALMs is their strong zero-shot learning ability \cite{wang2023neural, ji2024mobilespeech, rag_audio}, enabling them to effectively generalize to new and unseen tasks without explicit task-specific training. This capability is particularly valuable in domains with limited or rapidly evolving labeled data, such as audio spoofing detection.

\begin{figure}[h!]
    \centering
    \includegraphics[width=\columnwidth]{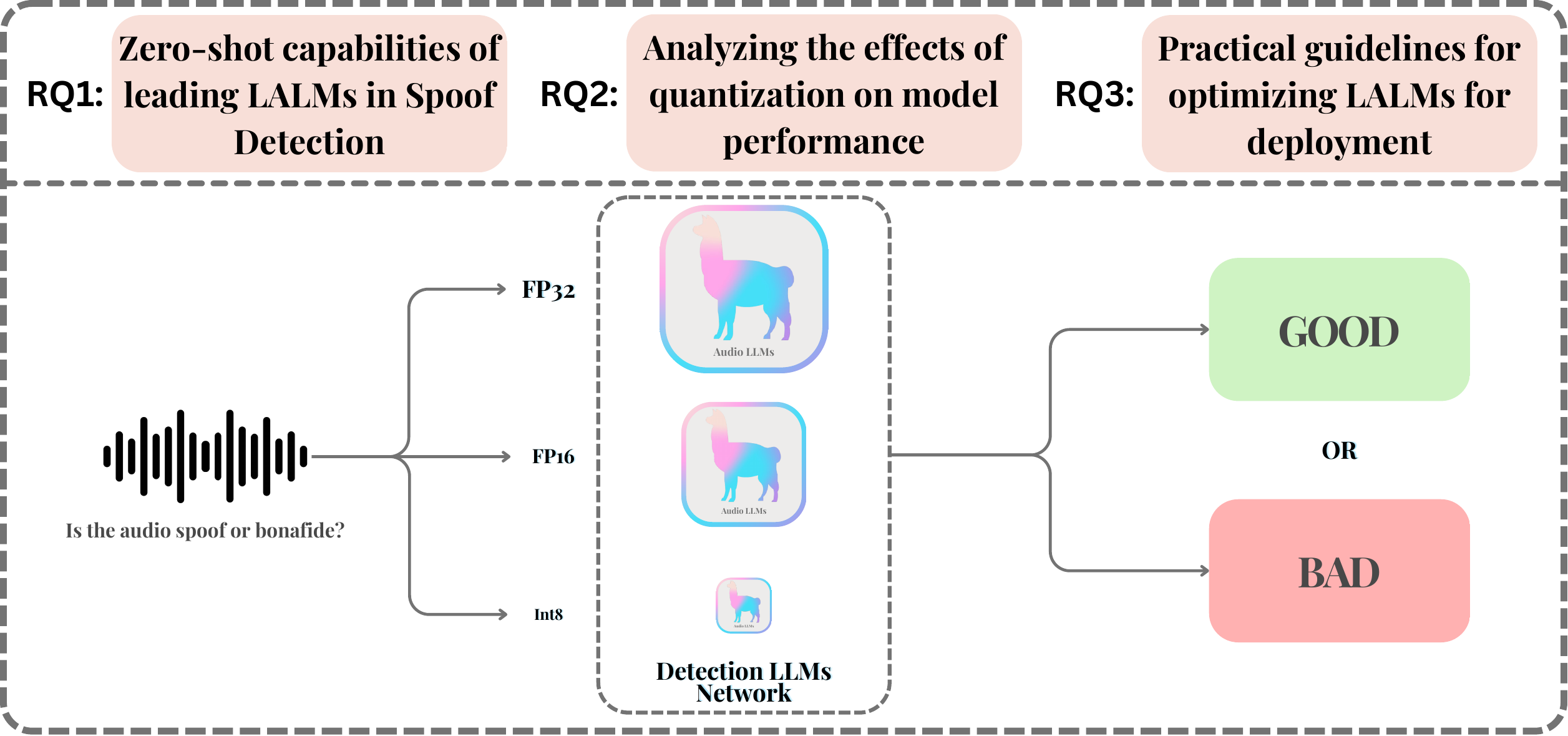}
    \caption{Overview of our evaluation framework addressing three research questions (RQs). RQ1 evaluates the zero-shot spoof detection capabilities of prominent LALMs. RQ2 investigates the impact of quantization precision (FP32, FP16, INT8) on model performance and computational efficiency. RQ3 offers practical guidelines for optimizing LALMs for real-world deployment, categorizing model performance as either ‘GOOD’ or ‘BAD’ based on robustness and efficiency.}
    \label{fig:zero_shot_performance}
    \vspace{-10pt}
\end{figure}

Despite their powerful capabilities, the practical deployment of LALMs is challenging due to substantial computational resource requirements, especially on devices with limited memory and processing power. To address these challenges, quantization has emerged as the essential optimization strategy. Quantization involves reducing the numerical precision of model parameters from higher precision formats, such as 32-bit floating-point, to lower precision formats, such as 16-bit floating-point (FP16) or 8-bit integers (INT8) \cite{quant2021}. This precision reduction significantly decreases memory usage and boosts inference speeds, making it feasible to deploy sophisticated audio-language models in real-time and resource-constrained environments. However, quantization is not without trade-offs, and its impact on the performance of audio tasks such as zero-shot spoof detection, remains largely unexplored.

Audio spoofing detection is an important domain where the potential of LALMs and the challenges of their practical deployment intersect. Rapid advancements in audio synthesis technologies, such as text-to-speech (TTS) and voice conversion (VC), have enabled the creation of highly realistic synthetic audio. These spoofed audio samples pose substantial threats to speaker verification systems, voice-controlled devices, and forensic audio analysis \cite{DBLP:conf/icb/RanjanVS23, ranjan_faking_fluent, ranjan_context_encoded}. Traditional spoof detection methods primarily rely on labeled datasets that rapidly become obsolete due to the continuous evolution of spoofing techniques. Consequently, these conventional approaches struggle to maintain effectiveness against emerging and sophisticated attacks.

In this context, zero-shot learning offers a promising alternative by leveraging pretrained audio-language models to detect spoofed audio samples without requiring labeled examples of new or emerging spoofing techniques. However, the effectiveness and robustness of zero-shot spoof detection capabilities under different quantization conditions remain inadequately studied. Understanding the interplay between zero-shot learning and quantization is crucial for deploying robust audio models in resource-constrained real-world scenarios.


This study addresses this critical research gap by evaluating the zero-shot spoof detection performance of five state-of-the-art LALMs, GAMA, LTU-AS, MERaLiON, Qwen-Audio, and SALMONN, across different quantization precisions (16-bit and 8-bit). Through detailed experiments and analysis, we explore how these models perform in zero-shot detection scenarios and explicitly quantify the effects of quantization on their accuracy and reliability. Our key contributions are:

\begin{itemize}
    \item Conduct the first comprehensive analysis of zero-shot spoof detection capabilities of quantized LALMs, providing foundational insights into inherent strengths and limitations.
    \item Deliver empirical insights into trade-offs among quantization precision, model accuracy, and computational efficiency, enabling informed decision-making for practical deployments.
    \item Present practical guidelines and recommendations for optimizing LALMs, ensuring robust spoof detection performance within realistic computational constraints.
\end{itemize}

\begin{figure*}[]
    \centering
    \includegraphics[width=0.95\linewidth]{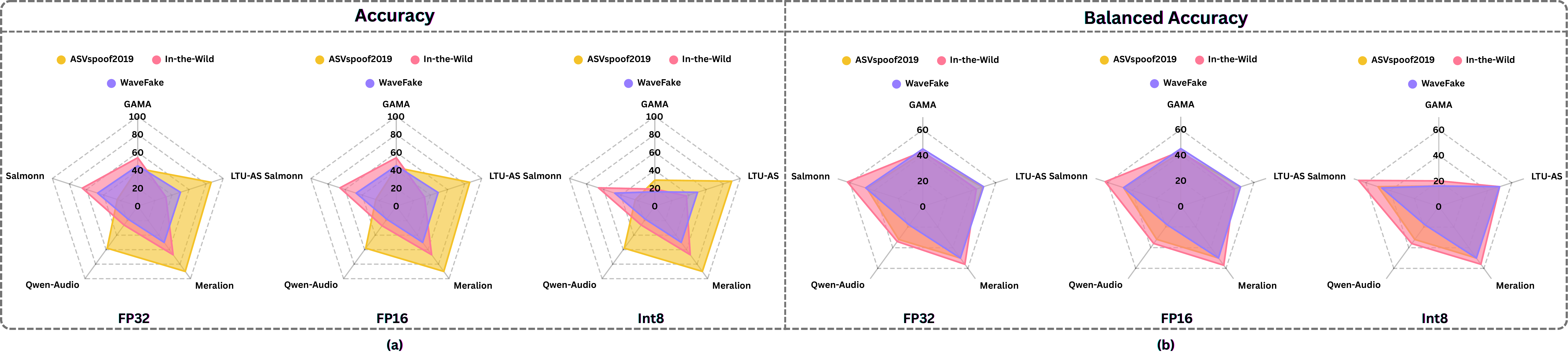}
    \vspace{-7pt}
    \caption{Illustrate \textbf{Accuracy (part a)} and \textbf{Balanced Accuracy (part b)} for five Large Audio Language Models (GAMA, LTU-AS, MERaLiON, Qwen-Audio, and SALMONN) across three datasets: ASVspoof2019 (yellow), In-the-Wild (pink), and WaveFake (purple). The results are presented for three quantization levels: FP32 (left), FP16 (center), and Int8 (right). Each plot highlights the models' ability to detect audio spoofing attacks under varying precision settings, showcasing differences in robustness and generalization capabilities across datasets and quantization strategies.}
    \label{fig:accuracy_plot}
    \vspace{-5pt}
\end{figure*}

\section{Formulation and Experimental Setup}

Consider an audio sample represented as a sequence of acoustic frames \( X = \{x_1, x_2, \dots, x_T\} \), where each frame \( x_i \in \mathbb{R}^d \), \( T \) is the total number of frames, and \( d \) is the dimensionality of each frame. Let \( f_\theta \) denote a pretrained LALM parameterized by weights \( \theta \). Given an audio sample ${X}$ and a prompt $p$, the LALM outputs a probability distribution over the vocabulary $\mathcal{V}$: $\mathbf{P} = f_\theta({X}, p) \in \Delta^{|\mathcal{V}|-1}$. The binary classification decision is based on comparing the probabilities of the target classification tokens, assigning the class $y = \text{spoof} \text{ if } P(t_{\text{spoof}} \mid X, p) > P(t_{\text{bonafide}} \mid X, p), \text{ else bonafide}$, where \( t_{\text{spoof}}, t_{\text{bonafide}} \in \mathcal{V} \) are the vocabulary tokens corresponding to the class labels.


In the zero-shot spoof detection scenario, \( f_\theta \) leverages general audio representations learned during pretraining, without explicit fine-tuning on spoof-specific data. Model quantization reduces the precision of parameters, defined by $\theta_{\text{quant}} = \mathcal{Q}_p(\theta)$, where $p \in \{32, 16, 8\}$ represents precision levels (FP32, FP16, INT8 respectively), and $\mathcal{Q}_p$ is the quantization function converting high-precision parameters into lower-precision formats.
This study systematically explores three key research questions: \textbf{(i) RQ1 - Zero-shot spoof detection}, which investigates the intrinsic capability of pretrained LALMs to identify spoofed audio without explicit fine-tuning; \textbf{(ii) RQ2 - Impact of quantization}, examining how varying quantization precision (FP32, FP16, INT8) affects the performance of LALMs; and \textbf{(iii) RQ3 - Deployment optimization}, aimed at providing practical guidelines for selecting optimal quantization precision that balances detection performance with computational efficiency.

\noindent \textbf{Experimental Framework and Baseline Models:}
To rigorously assess the zero-shot spoof detection capabilities, robustness to quantization, and practical deployment considerations of LALMs, we developed a comprehensive experimental framework. We evaluated five representative state-of-the-art pretrained LALMs selected for their architectural diversity and proven effectiveness in audio-related tasks. Specifically, \textbf{GAMA \cite{ghosh-etal-2024-gama}} integrates an Audio Q-Former, which generalizes audio semantics, with multi-layer Audio Spectrogram Transformer (AST) modules capturing detailed audio characteristics, leveraging the pretrained Llama model \cite{touvron2023llama} for advanced audio understanding. \textbf{LTU-AS \cite{gong_ltuas}} combines Whisper's robust encoder-decoder architecture \cite{radford2022whisper} with a specialized time- and layer-wise transformer (TLTR) to jointly process speech transcription and audio event detection, further refining these representations through integration with Llama. \textbf{MERaLiON \cite{he2024meralionaudiollmtechnicalreport}}, specifically tailored for multilingual environments, combines a localized version of Whisper-large-v2 with SEA-LION V3 \cite{sea_lion_2024}, and incorporates an adaptor module explicitly designed to enhance audio-text alignment and multilingual capability. \textbf{Qwen-Audio \cite{Qwen-Audio}}, developed by Alibaba, fuses Whisper-base’s audio processing capabilities with a transformer decoder, enabling versatile audio-text alignment applicable across various audio forms including speech, music, and environmental sounds. Finally, \textbf{SALMONN \cite{tang2024salmonn}} integrates Whisper and BEATs audio encoders through a window-level Q-Former combined with a Vicuna-based \cite{vicuna2023} LLM, providing unified and sophisticated multimodal processing across speech, audio events, and music.

\begin{table}[]
\caption{Distribution of spoofed and bonafide samples across ASVspoof2019, In-the-Wild, and WaveFake datasets.}
\vspace{-5pt}
\resizebox{\columnwidth}{!}{%
\begin{tabular}{|l|c|c|c|}
\hline
Dataset      & \multicolumn{1}{l|}{Spoof} & \multicolumn{1}{l|}{Bonafide} & \multicolumn{1}{l|}{Total Samples} \\ \hline
ASVspoof2019 & 63882                      & 7355                          & 71237                              \\ \hline
In-the-Wild  & 11816                      & 19963                         & 31779                              \\ \hline
WaveFake     & 13100                      & 13100                         & 26200                              \\ \hline
\end{tabular}
}
\label{table:datastats}
\vspace{-15pt}
\end{table}

Our experiments utilize multiple diverse datasets containing bonafide and synthetically spoofed audio samples that were created using various synthesis methods. We specifically selected datasets that reflect distinct class distributions, thereby allowing comprehensive evaluation under realistic scenarios: ASVspoof2019 \cite{nautsch2021asvspoof} is significantly skewed toward spoof samples, making it challenging for models to correctly identify genuine audio; In-the-Wild \cite{muller2022does} predominantly comprises bonafide samples, testing the models’ sensitivity and specificity in a practical context; and WaveFake \cite{frank2021wavefakedatasetfacilitate} presents a balanced dataset equally composed of spoofed and bonafide samples, providing a controlled setting for evaluating model generalization and robustness. Table \ref{table:datastats} summarizes key dataset characteristics. This carefully chosen suite of datasets enables insightful analysis and enhances the practical relevance of our findings for effectively deploying robust LALMs in real-world applications.

\noindent \textbf{Evaluation Metrics and Implementation Details:}
To thoroughly evaluate model performance, we employ a suite of complementary metrics: F1-Score, Accuracy, Balanced Accuracy, Matthews Correlation Coefficient (MCC), and classwise accuracy. The F1-Score provides a balanced measure of precision and recall, particularly valuable in handling class imbalance. Accuracy captures the overall prediction correctness, while Balanced Accuracy averages recall across classes, making it highly effective for imbalanced scenarios. MCC offers a comprehensive assessment of classification quality by considering all confusion matrix categories. Classwise accuracy delivers detailed insights into model behavior across individual classes. All experiments were executed using an NVIDIA A100 GPU, leveraging mixed-precision inferencing to enhance computational efficiency and accelerate matrix computations.

\begin{table*}[h!]
\caption{Comparison of five Large Audio Language Models evaluated on the different datasets. Each row represents a specific metric: F1-Score, Accuracy, Balanced Accuracy, Matthews Correlation Coefficient (MCC), (Spoof or Bonafide) class-wise Accuracy. The columns list the performance of each model under three quantization levels (FP32, FP16, INT8), with values separated by slashes.}
\vspace{-7pt}

\resizebox{1\linewidth}{!}{
\begin{tabular}{|lccccc|}
\hline
\multicolumn{6}{|c|}{\textbf{ASVspoof2019 (FP32 / FP16 / INT8)}}                                                                                                                                                                                                        \\ \hline
\multicolumn{1}{|l|}{\textbf{Metrics / Model}}   & \multicolumn{1}{c|}{\textbf{GAMA}}            & \multicolumn{1}{c|}{\textbf{LTU-AS}}         & \multicolumn{1}{c|}{\textbf{Meralion}}     & \multicolumn{1}{c|}{\textbf{Qwen}}            & \textbf{Salmonn}          \\ \hline
\multicolumn{1}{|l|}{\textbf{F1-Score}}          & \multicolumn{1}{c|}{0.570 / 0.580 / 0.454}    & \multicolumn{1}{c|}{0.925 / 0.925 / 0.946}   & \multicolumn{1}{c|}{0.946 / 0.946 / 0.946} & \multicolumn{1}{c|}{0.734 / 0.733 / 0.731}    & 0.297 / 0.293 / 0.265    \\ \hline
\multicolumn{1}{|l|}{\textbf{Accuracy}}          & \multicolumn{1}{c|}{0.424 / 0.433 / 0.293}    & \multicolumn{1}{c|}{0.861 / 0.861 / 0.897}   & \multicolumn{1}{c|}{0.897 / 0.897 / 0.897} & \multicolumn{1}{c|}{0.579 / 0.578 / 0.576}    & 0.237 / 0.237 / 0.226    \\ \hline
\multicolumn{1}{|l|}{\textbf{Balanced Accuracy}} & \multicolumn{1}{c|}{0.415 / 0.422 / 0.164}    & \multicolumn{1}{c|}{0.488 / 0.489 / 0.499}   & \multicolumn{1}{c|}{0.500 / 0.500 / 0.500} & \multicolumn{1}{c|}{0.323 / 0.323 / 0.321}    & 0.458 / 0.467 / 0.496    \\ \hline
\multicolumn{1}{|l|}{\textbf{MCC}}               & \multicolumn{1}{c|}{-0.104 / -0.096 / -0.419} & \multicolumn{1}{c|}{-0.039 / -0.035 / 0.014} & \multicolumn{1}{c|}{0.000 / 0.000 / 0.000} & \multicolumn{1}{c|}{-0.231 / -0.232 / -0.233} & -0.065 / -0.053 / -0.007 \\ \hline
\multicolumn{1}{|l|}{\textbf{Spoof Accuracy}}    & \multicolumn{1}{c|}{0.426 / 0.436 / 0.327}    & \multicolumn{1}{c|}{0.958 / 0.957 / 1.000}   & \multicolumn{1}{c|}{1.000 / 1.000 / 1.000} & \multicolumn{1}{c|}{0.646 / 0.645 / 0.643}    & 0.179 / 0.177 / 0.156    \\ \hline
\multicolumn{1}{|l|}{\textbf{Bonafide Accuracy}} & \multicolumn{1}{c|}{0.405 / 0.407 / 0.000}    & \multicolumn{1}{c|}{0.017 / 0.020 / 0.001}   & \multicolumn{1}{c|}{0.000 / 0.000 / 0.000} & \multicolumn{1}{c|}{0.000 / 0.000 / 0.000}    & 0.737 / 0.756 / 0.836    \\ \hline
\end{tabular}
}
\label{table:performance}
\end{table*}

\begin{table*}[h!]
\resizebox{1\linewidth}{!}{

\begin{tabular}{|lccccc|}
\hline
\multicolumn{6}{|c|}{\textbf{In-the-Wild (FP32 / FP16 / INT8)}}                                                                                                                                                                                                     \\ \hline
\multicolumn{1}{|l|}{\textbf{Metrics / Model}}   & \multicolumn{1}{c|}{\textbf{GAMA}}            & \multicolumn{1}{c|}{\textbf{LTU-AS}}         & \multicolumn{1}{c|}{\textbf{Meralion}}     & \multicolumn{1}{c|}{\textbf{Qwen}}           & \textbf{Salmonn}       \\ \hline
\multicolumn{1}{|l|}{\textbf{F1-Score}}          & \multicolumn{1}{c|}{0.014 / 0.014 / 0.177}    & \multicolumn{1}{c|}{0.489 / 0.490 / 0.542}   & \multicolumn{1}{c|}{0.247 / 0.254 / 0.238} & \multicolumn{1}{c|}{0.408 / 0.425 / 0.418}   & 0.510 / 0.507 / 0.592 \\ \hline
\multicolumn{1}{|l|}{\textbf{Accuracy}}          & \multicolumn{1}{c|}{0.536 / 0.542 / 0.193}    & \multicolumn{1}{c|}{0.329 / 0.330 / 0.372}   & \multicolumn{1}{c|}{0.670 / 0.672 / 0.668} & \multicolumn{1}{c|}{0.256 / 0.270 / 0.264}   & 0.648 / 0.657 / 0.662 \\ \hline
\multicolumn{1}{|l|}{\textbf{Balanced Accuracy}} & \multicolumn{1}{c|}{0.429 / 0.433 / 0.201}    & \multicolumn{1}{c|}{0.438 / 0.490 / 0.500}   & \multicolumn{1}{c|}{0.563 / 0.565 / 0.560} & \multicolumn{1}{c|}{0.344 / 0.363 / 0.355}   & 0.616 / 0.620 / 0.661 \\ \hline
\multicolumn{1}{|l|}{\textbf{MCC}}               & \multicolumn{1}{c|}{-0.231 / -0.223 / -0.591} & \multicolumn{1}{c|}{-0.254 / -0.250 / 0.000} & \multicolumn{1}{c|}{0.246 / 0.250 / 0.238} & \multicolumn{1}{c|}{-0.254 / -0.250 / 0.000} & 0.236 / 0.313 / 0.247 \\ \hline
\multicolumn{1}{|l|}{\textbf{Spoof Accuracy}}    & \multicolumn{1}{c|}{0.009 / 0.009 / 0.233}    & \multicolumn{1}{c|}{0.864 / 0.866 / 1.000}   & \multicolumn{1}{c|}{0.146 / 0.150 / 0.140} & \multicolumn{1}{c|}{0.864 / 0.866 / 1.000}   & 0.493 / 0.475 / 0.659 \\ \hline
\multicolumn{1}{|l|}{\textbf{Bonafide Accuracy}} & \multicolumn{1}{c|}{0.849 / 0.170 / 0.858}    & \multicolumn{1}{c|}{0.012 / 0.013 / 0.000}   & \multicolumn{1}{c|}{0.981 / 0.980 / 0.981} & \multicolumn{1}{c|}{0.012 / 0.013 / 0.000}   & 0.740 / 0.764 / 0.664 \\ \hline
\end{tabular}
}
\end{table*}

\section{Results and Analysis}
This section presents an in-depth analysis addressing our three primary research questions using five state-of-the-art LALMs across multiple datasets and precision levels.


\begin{table*}[h!]

\resizebox{1\linewidth}{!}{
\begin{tabular}{|lccccc|}
\hline
\multicolumn{6}{|c|}{\textbf{WaveFake (FP32 / FP16 / INT8)}}                                                                                                                                                                                                            \\ \hline
\multicolumn{1}{|l|}{\textbf{Metrics / Model}}            & \multicolumn{1}{c|}{\textbf{GAMA}}            & \multicolumn{1}{c|}{\textbf{LTU-AS}}         & \multicolumn{1}{c|}{\textbf{Meralion}}     & \multicolumn{1}{c|}{\textbf{Qwen}}            & \textbf{Salmonn}          \\ \hline
\multicolumn{1}{|l|}{\textbf{F1-Score}}          & \multicolumn{1}{c|}{0.511 / 0.512 / 0.280}    & \multicolumn{1}{c|}{0.661 / 0.660 / 0.667}   & \multicolumn{1}{c|}{0.667 / 0.667 / 0.667} & \multicolumn{1}{c|}{0.300 / 0.301 / 0.300}    & 0.456 / 0.456 / 0.438    \\ \hline
\multicolumn{1}{|l|}{\textbf{Accuracy}}          & \multicolumn{1}{c|}{0.446 / 0.445 / 0.163}    & \multicolumn{1}{c|}{0.495 / 0.494 / 0.500}   & \multicolumn{1}{c|}{0.500 / 0.500 / 0.500} & \multicolumn{1}{c|}{0.177 / 0.177 / 0.176}    & 0.469 / 0.470 / 0.470    \\ \hline
\multicolumn{1}{|l|}{\textbf{Balanced Accuracy}} & \multicolumn{1}{c|}{0.446 / 0.445 / 0.163}    & \multicolumn{1}{c|}{0.495 / 0.494 / 0.500}   & \multicolumn{1}{c|}{0.500 / 0.500 / 0.500} & \multicolumn{1}{c|}{0.177 / 0.177 / 0.176}    & 0.469 / 0.470 / 0.470    \\ \hline
\multicolumn{1}{|l|}{\textbf{MCC}}               & \multicolumn{1}{c|}{-0.113 / -0.114 / -0.713} & \multicolumn{1}{c|}{-0.039 / -0.048 / 0.010} & \multicolumn{1}{c|}{0.000 / 0.000 / 0.000} & \multicolumn{1}{c|}{-0.691 / -0.691 / -0.692} & -0.061 / -0.061 / -0.060 \\ \hline
\multicolumn{1}{|l|}{\textbf{Spoof Accuracy}}    & \multicolumn{1}{c|}{0.580 / 0.582 / 0.325}    & \multicolumn{1}{c|}{0.982 / 0.981 / 1.000}   & \multicolumn{1}{c|}{1.000 / 1.000 / 1.000} & \multicolumn{1}{c|}{0.354 / 0.354 / 0.353}    & 0.445 / 0.445 / 0.414    \\ \hline
\multicolumn{1}{|l|}{\textbf{Bonafide Accuracy}} & \multicolumn{1}{c|}{0.311 / 0.309 / 0.000}    & \multicolumn{1}{c|}{0.009 / 0.008 / 0.000}   & \multicolumn{1}{c|}{0.000 / 0.000 / 0.000} & \multicolumn{1}{c|}{0.000 / 0.000 / 0.000}    & 0.494 / 0.494 / 0.527    \\ \hline
\end{tabular}
}
\vspace{-8pt}
\end{table*}

\noindent \textbf{RQ1 - Zero-Shot Detection Capabilities:}
We evaluated the zero-shot detection capabilities of GAMA, LTU-AS, MERaLiON, Qwen-Audio, and SALMONN across three distinct datasets: ASVspoof2019, In-the-Wild, and WaveFake. Initial results on the ASVspoof2019 indicated promising spoof detection performance, with MERaLiON and LTU-AS achieving high F1-scores of 0.946 and 0.925, respectively, at FP32 precision. However, deeper examination of balanced accuracy and class-wise accuracy metrics highlights critical issues. We observed a severe bias toward classifying most or all inputs as spoofed. This bias artificially inflated spoof detection accuracy while drastically reducing the ability to detect bonafide samples. As a result, the overall accuracy metrics were misleading and resembled random predictions.

Further dataset-wise analysis provides deeper insights into the distinct behaviors of each model. The spoof-heavy ASVspoof2019 dataset exacerbated model biases, causing robust models such as MERaLiON and LTU-AS to frequently misclassify genuine audio samples as spoofed. The balanced WaveFake dataset offered a clearer assessment of generalization capabilities, but here too, models exhibited substantial biases. The predominantly bonafide In-the-Wild dataset highlighted LALMs' difficulties in accurately detecting genuine audio, emphasizing fundamental limitations in these architectures.

Quantitative evaluations further emphasized these concerns. MERaLiON demonstrated exceptional stability across quantization levels, consistently achieving an identical F1-score (0.946) and overall accuracy (0.897). However, its perfect spoof detection (accuracy of 1.000) came at the complete expense of bonafide detection (accuracy of 0.000). LTU-AS displayed similarly robust spoof detection, with an anomalous improvement in spoof accuracy under INT8 precision (from ~0.958 to 1.000). This anomaly was not indicative of improved performance but rather a reinforcement of the spoof-classification bias. GAMA showed heightened sensitivity to quantization, with its balanced accuracy deteriorating sharply from 0.415 at FP32 precision to 0.164 under INT8 precision, indicating potential vulnerabilities.

Qwen-Audio maintained moderate yet consistent spoof detection performance (F1-scores around 0.732) across precision levels but entirely failed to detect bonafide samples. SALMONN uniquely presented relatively strong bonafide detection capabilities (0.737–0.836 accuracy), despite generally poor overall metrics (F1-score ranging from 0.265 to 0.297). Across all models, MCC values remained consistently low or negative, emphasizing severe prediction bias and inadequate balanced performance. These findings highlight a troubling trend: current LALMs are strongly biased towards spoof detection at the expense of accurate bonafide identification, reflecting critical limitations that must be addressed through substantial architectural refinements to achieve reliable, balanced spoof detection suitable for real-world applications.

\noindent\textbf{RQ2 - Effects of Quantization on Model Performance:} Evaluating quantization at FP32, FP16, and INT8 precision levels showed performance trade-offs, highlighting the varying sensitivity of LALMs to reduced numerical precision. Generally, lowering precision negatively impacted model accuracy and reliability. However, we observed instances where INT8 quantization seemingly improved certain metrics, notably spoof detection accuracy. Upon deeper inspection, we identified these apparent improvements as statistical artifacts resulting from increased classification bias toward the spoof class. Specifically, INT8-quantized models excessively favored spoof predictions, artificially inflating overall accuracy while severely degrading balanced accuracy and genuine audio detection capability


A practically significant finding was the minimal performance difference between FP32 and FP16 across most models. Models such as MERaLiON and LTU-AS demonstrated almost identical performance at both FP32 and FP16 precision levels, suggesting FP16 quantization as an optimal practical choice. Specifically, FP16 precision can substantially reduce memory usage, approximately halving the required computational resources, without negatively impacting genuine detection performance. This finding positions FP16 quantization as an efficient strategy for deploying LALMs in real-world, resource-constrained scenarios.

GAMA, on the other hand, exhibited pronounced sensitivity to quantization, with a dramatic decline in balanced accuracy from FP32 to INT8 precision, indicating that certain architectural elements may significantly influence model resilience to precision reduction. In contrast, MERaLiON maintained consistency across all precision levels, potentially attributable to its adaptive audio-text alignment modules and multilingual training strategy. These modules likely enhance the robustness of internal representations, helping mitigate performance degradation under reduced precision conditions. Similarly, LTU-AS’s performance stability at FP16 precision suggests that integrating multiple transformer layers and multimodal embedding strategies can provide quantization resilience.

\begin{table}[]
\caption{Comparison of memory usage (GB) and inference time (seconds per 100 samples) at different quantization precisions.}
\vspace{-5pt}
\label{table:mem-time}
\resizebox{1\linewidth}{!}{
\begin{tabular}{|l|cc|cc|cc|}
\hline
\multirow{2}{*}{\textbf{Model / Factor}} & \multicolumn{2}{c|}{\textbf{FP32}}                  & \multicolumn{2}{c|}{\textbf{FP16}}                  & \multicolumn{2}{c|}{\textbf{INT8}}                  \\ \cline{2-7} 
                                         & \multicolumn{1}{c|}{\textbf{Memory}} & \textbf{Time} & \multicolumn{1}{c|}{\textbf{Memory}} & \textbf{Time} & \multicolumn{1}{c|}{\textbf{Memory}} & \textbf{Time} \\ \hline
\textbf{GAMA}                            & \multicolumn{1}{c|}{26.19}           & 129.06        & \multicolumn{1}{c|}{13.12}           & 126.20        & \multicolumn{1}{c|}{6.94}            & 177.76        \\ \hline
\textbf{LTU-AS}                          & \multicolumn{1}{c|}{25.35}           & 98.67         & \multicolumn{1}{c|}{12.71}           & 91.43         & \multicolumn{1}{c|}{6.74}            & 150.49        \\ \hline
\textbf{Meralion}                        & \multicolumn{1}{c|}{37.05}           & 90.77         & \multicolumn{1}{c|}{18.49}           & 57.55         & \multicolumn{1}{c|}{10.24}           & 107.31        \\ \hline
\textbf{Qwen}                            & \multicolumn{1}{c|}{31.35}           & 52.51         & \multicolumn{1}{c|}{15.64}           & 50.06         & \multicolumn{1}{c|}{9.12}            & 112.87        \\ \hline
\textbf{Salmonn}                          & \multicolumn{1}{c|}{51.47}           & 1020.38       & \multicolumn{1}{c|}{27.34}           & 960.38        & \multicolumn{1}{c|}{16.08}           & 1140.80       \\ \hline
\end{tabular}
}
\vspace{-10pt}
\end{table}










As precision decreases from FP32 to INT8, memory consumption consistently drops across all models. Moving from FP32 to FP16 reduces both memory consumption and inference time for all models as shown in Table \ref{table:mem-time}, demonstrating clear efficiency gains. However, while INT8 quantization achieves the greatest memory reduction, it paradoxically increases inference time compared to FP16. This suggests that INT8 quantization, while highly effective for memory optimization, introduces computational overhead due to data type conversions and associated processing, negating its execution-speed benefits.

\noindent\textbf{RQ3 - Practical Guidelines for Deployment:}
In our experiments, none of the tested LALMs demonstrated sufficient balanced performance or reliability suitable for immediate real-world deployment in spoof detection scenarios. The prevalent issue of pronounced predictive bias toward spoof classification significantly undermines their practical applicability, highlighting an urgent need for architectural and methodological refinements. Nevertheless, our findings offer clear and actionable insights for practical deployment strategies. FP16 quantization consistently emerged as the most favorable precision level across evaluated models, delivering a near-optimal balance between computational efficiency and detection accuracy. The minimal performance difference between FP32 and FP16 precision levels observed for models such as MERaLiON and LTU-AS reinforces the practicality of deploying FP16 quantized LALMs. Specifically, FP16 precision substantially reduces computational and memory requirements, effectively halving resource usage, making it particularly suitable for edge computing and mobile application scenarios.

INT8 quantization presents a more complex and nuanced trade-off. While theoretically ideal for extremely resource-constrained environments, our analysis shows significant risks, as INT8 quantization notably exacerbates predictive biases, severely compromising balanced classification performance. Therefore, INT8 deployment thus requires cautious consideration and rigorous pre-deployment validation processes to ensure reliability and performance integrity in practical scenarios.


Finally, our results suggest clear directions for future research and practical deployment enhancements. Architectural features such as adaptive alignment modules, robust multimodal embeddings, and multilingual training appear critical for enhancing quantization resilience. Incorporating these elements could mitigate quantization-induced biases, substantially improving the reliability and performance of LALMs in resource-constrained environments. These insights provide a foundational roadmap for future efforts aimed at refining LALMs into truly effective, balanced, and robust models suitable for real-world audio spoof detection.

\section{Conclusion}
This study presents the first comprehensive evaluation of zero-shot spoof detection capabilities of quantized LALMs, uncovering significant biases toward spoof classification that undermine practical applicability. Despite promising initial spoof detection results, deeper analysis of balanced metrics revealed fundamental limitations across all evaluated models, emphasizing the urgent need for substantial architectural refinements. Our analysis elucidates clear performance-precision trade-offs, highlighting FP16 quantization as an optimal choice that effectively halves memory and computational requirements with negligible accuracy degradation compared to FP32. Conversely, INT8 quantization, despite theoretical benefits in extreme resource-constrained scenarios, significantly intensifies predictive biases, requiring cautious consideration supported by rigorous pre-deployment validation. Additionally, our results showcased an interesting trade-off: while INT8 quantization significantly reduces memory consumption, it unexpectedly increases inference time compared to FP16. This highlights the necessity of carefully evaluating quantization strategies not only in terms of memory but also computational speed. Future architectures should incorporate adaptive alignment modules, robust multimodal embeddings, and multilingual training for improved quantization resilience. These strategies provide foundational guidance for deploying reliable, efficient LALMs in spoof detection.

\vspace{-8pt}
\section{Acknowledgment}
This research is supported by a grant from the NSM, MeitY. The authors also gratefully acknowledge the support of IndiaAI and Meta through Srijan: Centre of Excellence for Generative AI.

\bibliographystyle{IEEEtran}
\bibliography{refs, wav}

\end{document}